\pdfoutput=1
\documentclass[10pt, a4paper, twocolumn, showabstract]{naverlabseurope}

\usepackage{multicol}
\usepackage{times}
\usepackage{latexsym}
\usepackage{lipsum}
\usepackage{array}
\usepackage{tabularx}
\usepackage[colorinlistoftodos]{todonotes}
\usepackage[T1]{fontenc}

\usepackage[utf8]{inputenc}

\usepackage{microtype}

\usepackage{inconsolata}

\usepackage{graphicx}
\usepackage{booktabs}
\usepackage{arydshln}
\usepackage{fontawesome5}
\usepackage{balance}
\usepackage{multirow}

\title{Reranking with Compressed Document Representation}

\correspondingauthor{  herve.dejean@naverlabs.com}

\authors{ Hervé Déjean, Stéphane Clinchant }
\affiliations{NAVER LABS Europe}
\contributions{}
\website{https://github.com/naver/bergen}
\websiteref{\href{https://github.com/naver/bergen}}

\begin{abstract}
Reranking, the process of refining the output of a first-stage retriever, is often considered computationally expensive, especially with Large Language Models. Borrowing from recent advances in document compression for RAG, we reduce the input size by compressing documents into fixed-size embedding representations. We then teach a reranker to use compressed inputs by distillation.
Although based on a billion-size model, our trained reranker using this compressed input can challenge smaller rerankers in terms of both effectiveness and efficiency, especially for long documents. Given that text compressors are still in their early development stages, we view this approach as promising.

\end{abstract}

\begin{document}
\maketitle

\section{Introduction}

 Information Retrieval (IR) is typically understood as a two-part process: a \textit{first-stage} designed to swiftly locate pertinent documents for a specific query, followed by a more costly refinement phase called \textit{reranking}. 
Initially performed with cross-encoder \cite{gao_rethink_2021,nogueira2020passage},  recent models have gradually been tested, including encoder-decoder \cite{pradeep2021expandomonoduo,RankT5,zhuang_setwise_2023} or decoder-only models, namely Large Language Models (LLM) \cite{ma_fine-tuning_2023,sun_is_2023}. 


    While these LLM-based rerankers exhibit remarkable capabilities (zero-shot scenarios, fewer relevance judgements), they are  much less efficient compared to traditional rerankers using cross-encoders
\cite{zhuang_setwise_2023,déjean2024comp} - which are already deemed as slower components in IR systems.
The efficiency of large language models remains an open challenge \cite{zhu2024largelanguagemodelsinformation}, with relatively few studies addressing this issue.
First, a late interaction architecture has been proposed for cross-encoder-based rerankers, but it comes at the cost in effectiveness \cite{gao-etal-2020-modularized,10.1145/3397271.3401093}.
Moreover, \citet{liu2025matryoshkarerankerflexiblereranking} presents an architecture which allows the user to customize the architecture of their reranker by configuring the depth and width  of LLM,  achieving a 2× speed-up compared to the full-scale model) with a sequence length of 1024.
\citet{gangi-reddy-etal-2024-first, zhuang_setwise_2023} propose to leverage the output logits of the first generated token to directly obtain a ranked ordering of the candidates, lowering the latency of listwise LLM rerankers by 50\%. 

\textbf{We propose a more radical approach by significantly reducing the reranker's input sequence length through compressed representation, thereby achieving a 10x speed-up\footnote{assuming an input document length of 512 tokens.} compared to the usual textual representation}. 

\begin{figure}[ht]
     \centering
     \includegraphics[width=\linewidth]{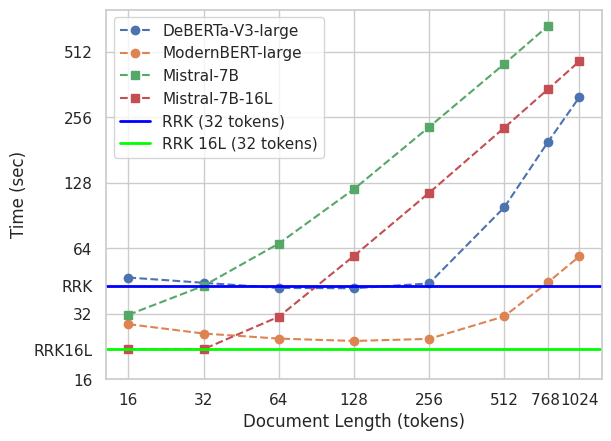}
     \caption{Reranking processing time according to the input length. Using compressed representation (RRK models) enables our reranker to maintain constant efficiency regardless of document length.
     }
     \label{fig:efficiency}
\end{figure}
First-stage dense retriever makes use of this idea of compressed input by representing a document with a single embedding. In order to  keep reranking effective, a richer representation is required. Colbert \cite{Santhanam2021ColBERTv2EA} can be considered  as  a reranker using a compact document representation.  
Recently, prompt compression methods have been developed to speed up LLMs facing long contexts, dialogue or in retrieval augmented generation \cite{icae}.
Recent works like \cite{cocom} and PISCO \cite{pisco} are able to learn compressed document representations in order to optimise a Retrieval Augmented Generation (RAG) system by effectively replacing retrieved texts by a much shorter compressed representation in the LLM prompt.
This raises an interesting question for retrieval: could a similar approach be used for reranking, relying solely on a compressed document embedding?

\textbf{Contribution:} We show in this paper that using such compressed representation in the context of reranking solves the efficiency problem of LLM-based rerankers while keeping comparable effectiveness. 
Additionally, using a small, constant input length through the use of compressed representation, enables reranking to maintain a constant efficiency regardless of document length (Figure~\ref{fig:efficiency}\footnote{The diagram illustrates the processing time for some models based on the input length. We simulate a set of 50 queries with 50 documents each, using a batch size of 256 on a A100 80G.}).


\section{Reranking Compressed Representation}
 We build on a recent soft \textit{offline} compression methods, PISCO, to learn our reranking models called \textbf{RRK}\footnote{RRK: compressed version of \textbf{R}e\textbf{R}an\textbf{K}er}. A RRK model consists in two models: a frozen PISCO compressor, mapping texts to embeddings, and a finetuned decoder, mapping query-document pairs to a reranking score. 
 
  First, each collection is compressed \textbf{offline} using the PISCO compressor: for each document $d_i$, a set of memory tokens $(m_1, \ldots, m_l)$ is appended, forming $(d_i; m_1, \ldots, m_l)$, which is fed to the compressor. The final $l$ hidden states of these memory tokens represent the document embeddings $\mathbf{c_i} = (c_i^s)_{s=1 \ldots l}$. In our case, we use $l=8$ memory tokens.
  To ensure consistent embeddings for reranking and RAG, the PISCO model remains frozen, \textbf{with only the decoder being fine-tuned using LoRA}.

At reranking stage, the compressed documents embeddings $\mathbf{c}$ are loaded and fed to the Decoder (Mistral-7B) finetuned for reranking.  To train the decoder, we use a mean squared error (MSE) loss to match the scores of a distilled teacher reranker, by adding a linear projection layer that projects the last layer representation of the end-of sequence token to a scalar.

We explore two RRK variants: one using all 32 layers of the Mistral-7B decoder, and a lighter version using only the first 16 layers for improved efficiency. In both cases, 8 memory tokens represent each document, and \textit{given a maximum query length of 24 tokens, the total input size remains fixed at 32 tokens}, thus  making the efficiency of the RRK model very competitive (see Figure~\ref{fig:efficiency}). Notably, our method shares conceptual similarities with {ColBERT}~\cite{Santhanam2021ColBERTv2EA}, which stores one embedding per token; however, \textbf{RRK} compresses each document into just 8 embeddings.
For the MSMARCO collection (8.8 million documents), the PISCO index results in a storage size of 270 GB, almost twice the size of the Colbert index (154 GB). 
However, in our case, this drawback is mitigated by the fact that  this index is also used during the generation step in the RAG setting.



\section{Training}
Training a high quality reranker depends on many factors such as labeled data with relevant documents and carefully chosen negatives from multiple retrievers, \cite{cao2024recentadvancestextembedding}. All these choices make comparison hard and reproducibility challenging due to varied and massive training sets.
But, a simpler and more reproducible solution is to rely on \textit{distillation} from existing rerankers. In fact, our goal is to assess whether it is possible to train rerankers from compressed representations and see whether LLM-based rerankers could be made faster. To do so, we take a state-of-the-art cross-encoder and distill it into LLMs, RRK models but also a more recent BERT baseline. Note that the distillation direction is inverted: from a small model to a larger one. 

In order to select our teacher,  we performed a set of evaluation using various first-stage and rerankers (see Appendix~\ref{appendix:teacher}). 
Based on those results, and aiming at building an efficient full RAG system, we choose the SPLADE-V3 sparse model  \cite{lassance2024spladev3newbaselinessplade}, faster than a dense model, and its  companion reranker Naver-DeBERTa \cite{lassance2023naverlabseuropesplade}.

For our training collection, we use the MS MARCO (passage) dataset \cite{msmarco}
The training collection, which consists of a set of queries and an appropriate document collection (without the need for relevance judgments), is processed using the selected first-stage retriever and reranker. For each query, we identify the top 50 documents produced by the reranker used as teacher. For the query set, we utilize the  0.5 million training queries, pairing each query with 8 documents randomly selected from the top 50 documents as provided by the retriever. We conduct training over 2 epochs, as additional epochs did not yield significant improvements. The finetuning takes 24h using 1 A100 GPU with a batch size of 8 and a learning rate of $1 \times 10^{-4}$.
While stronger training configurations could be developed, our focus is comparing compressed vs. textual representations, making this configuration suitable for a fair evaluation.


\begin{table*}[ht]
\centering
\tiny
\resizebox{1.0\textwidth}{!}{
\begin{tabular}{l |c|cc|cc|cc|cc|cc }
\toprule
 &
\textbf{SPLADE-V3} 
&  \multicolumn{2}{c}{\textbf{Naver-DeBERTa}}
&  \multicolumn{2}{c}{\textbf{ModernBERT}}
&  \multicolumn{2}{c}{\textbf{Mistral 7B}}
& \multicolumn{2}{c}{\textbf{RRK}}
& \multicolumn{2}{c}{\textbf{RRK 16 Layers}}
\\
&(retriever)& 256&512&256&512&256&512&256&512&256&512 \\
\midrule
\textbf{TREC} &\\
\midrule
DL 19 &72.3  & 77.6& 77.6 &76.3&76.3&\textbf{ 77.9}&\textbf{77.9}&77.1&77.1&77.1&77.1\\
DL 20 & 75.4 & 75.4&75.4 &76.7& 76.7&76.9&76.9&\textbf{77.0}&\textbf{77.0}&75.8&75.8\\
\midrule 
\textbf{BeIR} &\\
\midrule
TREC-COVID &74.8& 87.4&88.2&88.0& \textbf{89.0}    & 85.3& 85.6&86.4&86.2&85.9&86.5\\
NFCorpus   &35.7& 37.8&37.6&38.2& 38.1   & 37.9&    38.5 &38.0&\textbf{38.7}&38.3&38.9\\
NQ         &58.6&65.7 &65.7&65.9&66.0    &67.1 &\textbf{67.2}  &66.5&66.8    &65.1&65.2 \\
HotpotQA   &69.2&74.4 &74.5 &75.3&\textbf{75.4}     &74.1 &74.1  & 73.4 &73.4  & 71.4& 71.5\\
FIQA       &37.4&47.1 &47.8 &46.7&47.6     &47.0 &\textbf{48.2}  &46.7&47.1&45.1&46.1\\
Touché 2020-v2 &29.3  & 31.2&33.5 &32.9&\textbf{35.2}     &31.6 &31.6 &28.6&28.9&29.4&28.7 \\
Quora      & 81.4&84.3& 84.3&86.0&86.0    &86.4& 86.4  &\textbf{87.0}&\textbf{87.0}& \textbf{87.0}&\textbf{87.0}\\
DBPedia    &45.0&48.8 &48.8&50.1&50.1    &\textbf{50.5} &\textbf{50.5}  &49.4&49.4& 49.1&49.1\\
SCIDOCS    &15.8&19.3 &19.2&19.5&19.2     &\textbf{20.1} &19.2  &19.4&19.5&19.4&19.5 \\
FEVER      &79.6&83.5 &86.5 &85.76&\textbf{88.4}     &84.2 &86.9 & 83.4&84.2& 81.2&81.7 \\
Climate-FEVER &23.3 & 25.0 &27.4  &   23.2&25.3   &23.9 &2\textbf{7.5} &25.9&25.9& 25.8&25.9\\
SciFact    &71.0&76.2 &75.8&75.3&75.4    &75.4 &\textbf{77.3}  &75.2&76.4& 75.4&74.5 \\
\midrule
\textbf{AVG} & 51.8  &\textbf{56.6} &\textbf{57.4} &\textbf{57.2} &\textbf{57.9} & \textbf{57.0}&\textbf{57.8} & \textbf{56.6} &\textbf{56.9} & \textbf{56.1}&\textbf{56.2} \\
\midrule 
\textbf{LoTTe}  &  \\
\midrule
 pooled search  &53.3 &62.2&62.6 &62.1&62.5 &62.5&\textbf{62.9}& 62.4&62.6 & 61.4&61.1 \\
 pooled forum & 36.0& 45.8&46.4& 45.4&45.9 & 45.9& \textbf{46.4} &45.9&46.3 & 45.3&45.2\\ 
\bottomrule
\end{tabular}
}

\caption{
Evaluation ($nDCG@10 * 100$) of textual and compressed (RRK) rerankers . The top 50 of the SPLADE-v3 retrieved documents is used as input. 
}
\label{tab:reranker}
\end{table*}

\section{Evaluation}
For evaluation, we use traditional IR benchmarks TREC-DL 19/20  \cite{craswell2020overview,craswell2021overview}, BeIR \cite{Thakur2021BEIRAH}, and due to space constraints, we have only included the pooled subset of the LoTTe collection (forum and search) \cite{Santhanam2021ColBERTv2EA}. As mentioned in \cite{déjean2024comp, E2Rank}, we ignore the BeIR Arguana collection, as this collection aims at finding counter-arguments. As for training, the top 50 documents from the first-stage, SPLADE-v3,  are reranked. 
nDCG@10 is used as evaluation measure for all datasets. We provide for all models their results with a  maximum document length of 
both 256 and 512 tokens. 

In order to compare the use of compression with regard to the textual representation, we train with the same setting two models using textual input: a Mistral-7B as this model was used for the PISCO/RRK model, and ModernBERT-large \cite{warner2024smarterbetterfasterlonger}, a 'smaller, better, faster, longer'[\textit{sic}] bidirectional encoder, which shows very competitive results in terms of effectiveness, and efficiency (faster than DebertaV3). Appendix~\ref{appendix:reranker} shows that the ModernBERT-base version is not able to reach the teacher level.  

The PISCO compressor needs an access to the compressed representation during inference. This access amount for less than 10\% of the total reranking time. When discussing efficiency, we always refer to Figure~\ref{fig:efficiency}. 



\begin{table}[ht]
\tiny
\centering

\resizebox{0.5\textwidth}{!}{
\begin{tabular}{l|cccc}
\toprule
  BeIR/TREC-NEWS                     &       \multicolumn{4}{c}{\textbf{Document Length Input}} \\
                       &   \textbf{256}  & \textbf{512} & \textbf{768} & \textbf{1024} \\
\midrule                       
Naver-DeBERTa-v3-Large & 46.0            & 50.9       & 48.6          & 46.0   \\                       
ModernBERT-Large       &  47.4             & 50.5        & 51.2      & 51.5    \\
Mistral 7B            & 49.4              & 51.6        &52.3         &51.2\\
RRK                  &   46.5   & 51.2   & \textbf{53.1}       & 51.9\\

\end{tabular}
}

\caption{Reranking "Long" documents. Increasing the document length can enhance rerankers effectiveness without affecting 
 the efficiency of RRK models.}
\label{tab:long}
\end{table}

\section{Results}
The results are presented in Table~\ref{tab:reranker}. The first column provides information about the average number of words per query and document, which will be discussed later, and the second column evaluates our teacher model (see Table~\ref{appendix:models} for the first stage evaluation).


All models perform similarly, with differences of less than 1 point using nDCG@10 for the average score. Notably, the main differences often originate from a few specific datasets  (esp FEVER and Touché). Ignoring these datasets reduces the differences between ModernBERT and RRK by $50\%$. 
The recent ModernBERT performs very well and is very efficient, which shows that  the reranking task  can be performed with an encoder only architecture.  The Mistral model fine-tuned with textual representation achieves similar effectiveness than ModernBERT, but is simply 20 times slower. Compared with the RRK model, the differences among datasets are less than 1 point, except for FEVER and Touché. We can not explain yet this weakness of the RRK model with these two datasets. 

Let us first draw some some general observations: distillation is effective since all models replicate the teacher's results or improve it. Moreover, increasing input length improves results by 1 point. 
This might seem marginal, but the gain is more significant (>2 points) for collections with long documents, like  FEVER (full wikipedia pages). While models utilizing textual representation are directly affected by increased length, the RRK model is impacted only during offline compression, maintaining constant efficiency in online reranking. 

Secondly, we now discuss the key results on reranking with compressed representation:  
RRK, our model with a 32 token input length is up to 16 times faster than its equivalent using textual representation with almost the same effectiveness (-0.9 for BeIR, similar for LoTTe). Furthermore, RRK is also \textit{faster than its teacher} (Figure~\ref{fig:efficiency}), for the same effectiveness. Compared to our strong baseline, ModernBERT-Large, our RRK model is a bit less effective for BeIR (-1 pt), but achieves similar performance for LoTTe. Efficiency-wise,  it is half as fast (Figure~\ref{fig:efficiency}), but becomes slightly faster with long inputs (considering 768 tokens or more, see discussion below).

Regarding our RRK model with 16 layers, its efficiency is remarkable, being faster than Modern-BERT, though with  a reduced effectiveness (-2 points). Besides, we also attempted to train smaller PISCO models (like Qwen 3B or Llama 1B), but all attempts have been unsuccessful so far. Note that the RRK 16 layers is more effective and as fast as than the small ModernBERT-base model (Table~\ref{tab:models}).

Lastly, the use of the PISCO compressor, trained to compress 128 tokens into 8 memory tokens, shows promising results when applied to larger documents (512 tokens and beyond), as shown for the TREC-NEWS collection \cite{soboroff2018trec} in Table~\ref{tab:long}. This table demonstrates that the use of compressed representation is a natural setting for processing long documents: the RRK model achieves the best results with a 768 token input, while other models suffers losses in effectiveness, and/or a high increase of their efficiency (RRK is as fast as ModernBERT when the latter is fed with a 768 token input).
The decrease in performance observed with 1K tokens for most models may be due to the PISCO compressor being trained on relatively small documents (128 tokens or fewer). The reranker fine-tuning using this same maximum document length may also contribute to the decrease.

Overall, these results show that using compressed representations enables a 7B parameter model to run 10 times faster than its textual counterpart, and only half fast than a very recent 400M parameter model using all accumulated recipes for improving its effectiveness and efficiency. 
At this stage, we can see the glass as half empty or half full. A good choice is to distill your favorite reranker with ModernBERT. However the results we obtain with PISCO's compressed representation, trained for the generation part of a RAG system, are very appealing.


\section{Conclusion}
In this work, we introduce a novel approach to reranking by utilizing compressed document representations, significantly enhancing efficiency while maintaining similar levels of effectiveness. Our results demonstrate that employing compressed embeddings generated by the PISCO model—an off-the-shelf compressor model designed for Retrieval Augmented Generation—can achieve acceptable performance with better latency compared to traditional reranking methods, particularly when documents are sufficiently lengthy. This efficiency is achieved by simply reducing the input length for the model through the use of a compressed representation consisting of a small sequence of tokens.


\section*{Limitations}
First, the efficiency of RRK is mostly due to its tiny input length. This advantage  holds  as long as the query itself is short. Using datasets like \citet{BRIGHT}, where queries length is comparable to (BeIR) document length,  breaks this advantage and makes the RRK model slow. An evaluation with a query length of 48 tokens yields similar results for the BeIR collection except for the Scifact dataset, where the nDCG score rises from 75.1 to 76.4.

Secondly, it would be beneficial to employ smaller models instead of billion-sized ones as reranker. Unfortunately, our initial attempts to use smaller models, such as 1B parameter models, have not yet been successful (to say the least, they failed miserably). Using smaller models would lead to even better efficiency, and may reduce the index footprint (using smaller hidden dimensions).


Finally, the fact that the PISCO compressor as well as the RRK model have been trained with a maximum input length of 128 which certainly limits the effectiveness of the RRK model for long documents. 

\bibliography{main}
\bibliographystyle{acl_natbib}

\newpage
\appendix

\section{Selecting our Teacher}
\label{appendix:teacher}
All the next experiments have been generated using the BERGEN benchmark \cite{rau2024bergen} and are easily reproducible.
Table \ref{tab:whichreranker} presents a combination of publicly available first-stage retrievers and rerankers, including the traditional BM25 retriever (see Table~\ref{tab:models} in Appendix for more details about the models). Although it showcases a limited number of models, this table demonstrates how to select an appropriate teacher for the distillation process. Depending on the choice of your first-stage retriever, which may be influenced by efficiency constraints, certain available rerankers may be more suitable than others. 

Based on those results, MixBread seems like a viable option as a retriever. However, given that the Splade/DeBERTa combination offers similar performance to MixBread/MixBread one,  and \textbf{prioritizing efficiency}, we choose the sparse model for its greater speed than a dense one (comparable to  BM25 with an optimized implementation \cite{lassance2024twostepspladesimpleefficient,Bruch_2024} ). Additionally, both models have only been trained with the MSMARCO data, which makes comparison and further investigation easier. 
Therefore, in this paper, we choose SPLADE-V3 as the retriever and provide the top 50 documents it retrieves to the Naver-DeBERTa reranker. We use Naver-DeBERTa as our teacher model for distillation.

\begin{table*}[ht]

\centering

\resizebox{1.0\textwidth}{!}{
\begin{tabular}{l cccccc  ccc cccc cccc  }
\toprule
Retriever 
& \texttt{BM25}
& \texttt{SPLADE-v3}
& \texttt{Mix}
& \texttt{BGE}
& BM25 & BM25 & BM25
& SPLADE-v3 &MIXB & BGE

\\
        &&&&& +&+&+&+&+&+\\
Reranker&&&&& DeBERTa & mix &BGE & DeBERTa & mix &BGE \\
\midrule
\textbf{BeIR} &\\
\midrule    
Trec-covid & 0.59 & 0.75 & 0.76 & 0.75 & 0.82 & 0.81 & 0.73     & 0.85 & 0.87 & 0.79\\
SCIDOCS    & 0.15 & 0.16 & 0.23 & 0.22 & 0.19 & 0.19 & 0.17     & 0.19 & 0.20 & 0.18\\
NQ         & 0.31 & 0.59 & 0.56 & 0.55 & 0.54 & 0.53 & 0.53      & 0.66 & 0.64 & 0.50\\
SciFact    & 0.68 & 0.71 & 0.74 & 0.74 & 0.76 & 0.74 & 0.70     & 0.76 & 0.75 & 0.69\\
FIQA       & 0.24 & 0.37 & 0.45 & 0.44 & 0.40 & 0.37 & 0.34     & 0.45 & 0.44 & 0.39\\
QUORA      & 0.79 & 0.81 & 0.89 & 0.89 & 0.85 & 0.76 & 0.80     & 0.84 & 0.70 & 0.75\\
NFCorpus   & 0.34 & 0.36 & 0.38 & 0.37 & 0.38 & 0.39 & 0.33     & 0.38 & 0.40 & 0.31\\
HotpotQA   & 0.63 & 0.69 & 0.72 & 0.74 & 0.73 & 0.76 & 0.78     & 0.74 & 0.74 & 0.84\\
DPPedia    & 0.32 & 0.45 & 0.45 & 0.44 & 0.42 & 0.43 & 0.43     & 0.49 & 0.49 & 0.49\\
FEVER      & 0.65 & 0.80 & 0.86 & 0.84 & 0.80 & 0.78 & 0.87     & 0.78 & 0.81 & 0.92\\
Climate-FEVER & 0.17& 0.23&0.34&0.28&0.24&0.24&0.31&0.25&0.27& 0.37\\
Touché v2  & 0.29 &  0.29 & 0.23    &  0.23  & 0.34 & 0.35     & 0.31     & 0.31 & 0.24  &0.24 \\
\midrule
AVG &0.43& 0.52 & \textbf{0.55} &0.54&0.54& 0.54& 0.53 &\textbf{0.56}& \textbf{0.56}& 0.54 \\
\bottomrule
\end{tabular}
}

\caption{Which reranker should you use? Rerankers may strongly depend on the choice of first-stage retriever. NDCG@10. Rerankers consider the top 50 documents and a maximun document length of 256.}
\label{tab:whichreranker}
\end{table*}



\section{Rerankers}
\label{appendix:reranker}
Since we were only using a Mistral 7B backbone in this paper, it would be interesting to see whether other backbone or size could change our current results. We are currently only able to provide some results for a non compression setting in Table~\ref{tab:rerankernocompr}. We first see that the base version  of MordernBERT is not able to replicate the teacher results. Small LLMs as Llama3.2 1B is able to replicate them, but as mentioned, we were not able to train a PISCO using such backbone. For larger LLMs (7 billion parameters), no noticeable difference appears.

\begin{table*}[ht]
\centering

\resizebox{1.0\textwidth}{!}{
\begin{tabular}{l ccccccccccc }
\toprule
\textbf{}
& \textbf{SPLADE-V3}
&  \textbf{DeBERTa-v3}
&  \textbf{ModernBERT-base}
& \textbf{ModernBERT-large }
& \textbf{llama 1B} 
& \textbf{Mistral 7B} 
& \textbf{LLama 8B}   
& \textbf{Qwen 7B} 
\\
 & (Retriever)  & (Teacher)  \\
\midrule
\midrule
\textbf{BeIR} &\\
\midrule
TREC-COVID&0.75 &0.87&0.88  &0.88&0.86 &0.85&0.85 &0.85 \\
NFCorpus  & 0.36 &0.38 & 0.36 &0.38&0.37&0.38&0.30.38&0.39\\
NQ        &0.59&0.66& 0.62& 0.65 &0.66&0.67 &0.67&0.67\\
HotpotQA&0.69  &0.74 &0.72&0.74&0.75&0.74&0.74&0.75\\
FIQA & 0.37 &0.47&0.43 &0.47&0.45 &0.47&0.47 &0.47\\
Touché 2020-v2 & 0.29  & 0.31 &0.31&0.32&0.31&0.32&0.31&0.30\\
Quora &0.81 &0.84&0.83&0.86 &0.86&0.86& 0.87&0.85\\
DBPedia&0.45&0.49&0.48&0.49 &0.50&0.51& 0.51&0.51\\
SCIDOCS&0.16 &0.19&0.17&0.19 &0.20&0.20& 0.20&0.20\\
FEVER &0.80&0.83&0.88&0.85 &0.86&0.84 &0.85 &0.85\\
Climate-FEVER  &0.23 &0.25&0.21&0.23& 0.25&0.24& 0.26&0.25\\
SciFact  &0.71&0.76&0.63&0.76 &0.75&0.75& 0.76&0.77\\
\midrule
AVG & \textbf{0.52}&\textbf{0.56} & \textbf{0.54}&\textbf{ 0.57}  &\textbf{ 0.57 } & \textbf{ 0.57 }& \textbf{0.57}& \textbf{0.57} \\
\midrule
\bottomrule
\end{tabular}
}

\caption{
Evaluation (NDCG@10) of rerankers trained with\textbf{ no compression} using Naver-DeBERTa as teacher. The top 50 documents of the SPLADE-v3 retrieved documents is used as input. Maximum input length 256. 
}
\label{tab:rerankernocompr}
\end{table*}

\section{First-Stage Retriever and Rerankers}
\label{appendix:models}
Table~\ref{tab:models} provides the list of models used in this paper.

\begin{table*}[ht]
\centering

\resizebox{0.8\textwidth}{!}{
\begin{tabular}{lll}
\toprule
              & \textbf{References} & \textbf{Hugging Face model }\\
\midrule
\textbf{Off-the-shelf Retriever} & \\
SPLADE-v3 &\citet{lassance2024spladev3newbaselinessplade}& naver/splade-v3 \\
BGE &\citet{bge} &BAAI/bge-large-en-v1.5  \\
MIXBREAD  &\citet{mixbread} & mixedbread-ai/mxbai-embed-large-v1\\
\textbf{Off-the-self Reranker} &  \\
Naver-Deberta &\citet{lassance2023naverlabseuropesplade} & naver/trecdl22-crossencoder-debertav3 \\
MIXBREAD &\citet{mixbread}   &mixedbread-ai/mxbai-rerank-large-v1 \\
BGE & \citet{bge} & BAAI/bge-reranker-large  \\
\textbf{Trained Models (Reranker) } \\
Mistral & &mistralai/Mistral-7B-v0.3 \\
ModernBERT-large &\citet{warner2024smarterbetterfasterlonger} &answerdotai/ModernBERT-large \\
\textbf{Compressor} & \\
PISCO   &\citet{pisco} & naver/pisco-mistral \\  
\end{tabular}
}

\caption{
Models used in this article for retrieving, reranking and compressing.
}
\label{tab:models}
\end{table*}

\section{PISCO Architecture}
\begin{figure*}[ht]
     \centering
     \includegraphics{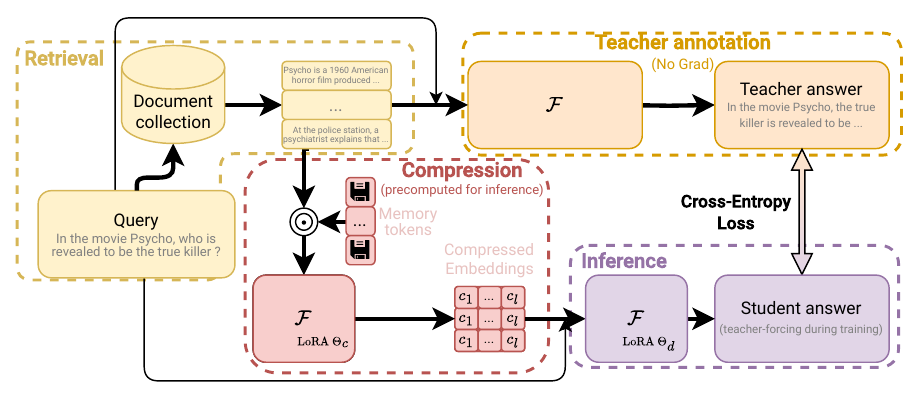}
     \caption{PISCO Architecture \cite{pisco}: The compression process utilizes a language model with LoRA adapters, appending memory tokens to each document to form embeddings, which control the compression rate through optimization. Decoding involves fine-tuning the decoder to adapt generation with compressed representations based on queries. The distillation objective employs Sequence-level Knowledge Distillation (SKD) to ensure models give consistent answers whether inputs are compressed or not.}
     \label{fig:pisco}
\end{figure*}

Figure~\ref{fig:pisco} shows how a PISCO model is trained. 

\end{document}